\documentclass[sn-mathphys-num]{sn-jnl}

\usepackage{amsmath,amssymb,amsfonts}%
\usepackage{amsthm}%
\usepackage{comment}
\usepackage{graphicx}%
\usepackage{multirow}%
\usepackage[latin1]{inputenc}
\usepackage{mathrsfs}%
\usepackage[title]{appendix}%
\usepackage{xcolor}%
\usepackage{textcomp}%
\usepackage{manyfoot}%
\usepackage{booktabs}%
\usepackage{algorithm}%
\usepackage{algorithmicx}%
\usepackage{algpseudocode}%
\usepackage{listings}%

\theoremstyle{thmstyleone}%
\theoremstyle{thmstyletwo}%

\theoremstyle{thmstylethree}%

\newcommand{\diff}{\mathrm{d}}

\begin{document}

\title[Minimal-length quantum field theory: a first-principle approach]{Minimal-length quantum field theory: a first-principle approach}

\author*[1]{\fnm{Pasquale} \sur{Bosso}}\email{pasquale.bosso@ino.cnr.it}

\affil*[1]{\orgname{CNR-INO, Istituto Nazionale di Ottica}, \orgaddress{\street{Via Campi Flegrei 34}, \city{Pozzuoli}, \postcode{I-80078}, \country{Italy}}}

\abstract{
  Phenomenological models of quantum gravity often consider the existence of some form of minimal length.
  This feature is commonly described in the context of quantum mechanics and using the corresponding formalism and techniques.
  Although few attempts at a quantum field-theoretical description of a minimal length has been proposed, they are rather the exception and there is no general agreement on the correct one.
  Here, using the quantum-mechanical model as a guidance, we propose a first-principle definition of a quantum field theory including a minimal length.
  Specifically, we propose a two-step procedure, by first describing the quantum-mechanical models as a classical field theory and subsequently quantizing it.
  We are thus able to provide a foundation for further exploration of the implications of a minimal length in quantum field theory.
}


\maketitle

\section{Introduction}\label{sec1}

A minimal measurable length, explicated in various different forms, is present in several approaches to quantum gravity \cite{Mead:1964zz,Padmanabhan:1987au,Konishi:1989wk,Ng:1993jb,Garay:1994en,Adler:1999bu,Scardigli:1999jh,Calmet:2004mp}.
From the phenomenological point of view, a relevant and interesting question concerns effects of such a minimal length on low-energy systems \cite{Ashtekar:2002sn,Amelino-Camelia:2008aez,Pesci:2018syy,Bosso:2018ufr,Lake:2018zeg}.
A possible strategy to introduce a minimal length, in the form of a minimal uncertainty in localization, is to modify the commutation relation between position and momentum, that is, the mathematical structure of quantum mechanics implying the uncertainty relation between these two quantities \cite{Kempf:1994su,Maggiore:1993rv,Das:2009hs,Pikovski:2011zk,Marin:2013pga,Bosso:2016ycv,Casadio:2020rsj,Segreto:2022clx}.
In what follows, when we will refer to a minimal length, we will indeed mean a minimal uncertainty in position, specifically.
Studying such deformed models can then be used as a probe for features in high-energy physics, aiming for an indirect observation of quantum gravitational effects.

Currently, quantum field theory serves as the indispensable theory to describe high-energy phenomena.
Therefore, it represents the necessary language for a phenomenological description of quantum gravitational effects.
Unfortunately, when it comes to models concerning the description and properties of a minimal length, quantum field theoretical models are rather the exception than the rule (see, \emph{e.g.}, \cite{Husain:2012im,Bosso:2020fos,Bosso:2021koi}.)
Furthermore, the few models present in the literature are usually inspired by the mathematical structure of the quantum mechanical counterpart, rather than by introducing a consistent construction that could lead to physical features associated with a minimal length.
In this work, we intend to solve this issue via a two-step procedure.
First, starting from the description of minimal-length quantum mechanics, we will rewrite it as a classical field theory, mimicking the way ordinary quantum mechanics can be described as a classical field theory, that is, in terms of a classical field (the wavefunction) whose equation of motion is the Schr\"odinger equation.
Along the way, studying the symmetry by spatial translations, time shifts, and complex phase shifts, we will find the corresponding conserved currents and densities.
Second, we will derive a (non-relativistic) quantum field theoretical formulation for a minimal length by quantizing the classical counterpart, thus effectively introducing a second quantization.
We already anticipate that this procedure will result in a model in which the commutation relation between fields is not deformed, contrarily to what is commonly assumed in the literature.

This work is organized as follows.
In Section \ref{sec:CFT_review}, we review the main features of quantum mechanics as a classical field theory.
In Section \ref{sec:MLQM_review}, we review the fundamental and necessary properties of a quantum-mechanical description of a minimal measurable length.
In Section \ref{sec:MLCFT}, we present the classical field-theoretical description for a minimal length.
In Section \ref{sec:symmetries}, we study various symmetries and find the corresponding conserved currents and densities.
In particular, we consider spatial translations (Sec. \ref{sec:MLCFT_spatial_translations}) and complex phase shifts (Sec. \ref{sec:MLCFT_phase_shifts}), while time shifts result to be trivial.
In Section \ref{sec:MLQFT}, we proceed with our plan by proposing a quantum field-theoretical description for a minimal length, including, as an example, the case of quartic interaction in Sec. \ref{sec:MLQFT_quartic}, thus explicitly showing the effects of such description.
Finally, in Section \ref{sec:conclusions}, we conclude summarizing the work.

A series of appendices are supplied with this paper.
Specifically:
Appendix \ref{apx:representations} describes the representations used in this work and the transforms relating them.
Appendix \ref{apx:current_MLQM} includes the necessary computations to obtain the conserved current and density associated with an arbitrary symmetry transformation.
Appendix \ref{apx:1D_current} specializes the results of the previous Appendix to the one-dimensional case.

\section{Quantum Mechanics as a Classical Field Theory}
\label{sec:CFT_review}

We start by reviewing the main feature of quantum mechanics spelled as a classical field theory \cite{Dickbook}.
Here and throughout the paper, we will consider spinless, scalar systems for simplicity.

Consider the following Lagrangian density
\begin{equation}
  \mathcal{L}
  = \frac{i \hbar}{2}\left(\Psi^{*} \partial_0 \Psi - \partial_0 \Psi^{*} \Psi\right) - \frac{\hbar^2}{2 m} \partial_i \Psi^{*} \partial^i \Psi-\Psi^{*} V \Psi,
  \label{eqn:ordinary_lagrangian}
\end{equation}
where $\Psi$ and $\Psi^*$ are two functions of the position vector $\vec{x}$ and are treated as two independent fields.
Here, we are using the notation $\partial_i$ to denote partial derivatives with respect to $x^i$, namely $\partial_i \equiv \frac{\partial}{\partial x^i}$, and $\partial_0$ to denote partial derivatives with respect to time, \emph{i.e.}, $\partial_0 = \frac{\partial}{\partial t}$.
The Lagrange equation for the field $\Psi^*$ is
\begin{equation}
  i \hbar \partial_0 \Psi
  = - \frac{\hbar^2}{2 m} \partial_i \partial^i \Psi + V \Psi,
  \label{eqn:ordinary_schrodinger}
\end{equation}
which corresponds to the Schrödinger equation for the field $\Psi$.

From the Lagrangian in Eq. \eqref{eqn:ordinary_lagrangian}, we find the momenta conjugate to the fields $\Psi$ and $\Psi^*$ to be, respectively,
\begin{align}
  \Pi
  &= \frac{\partial \mathcal{L}}{\partial (\partial_0 \Psi)}
  = \frac{i \hbar}{2} \Psi^{*},&
  \Pi^*
  &= \frac{\partial \mathcal{L}}{\partial (\partial_0 \Psi^*)}
  = - \frac{i \hbar}{2} \Psi.
  \label{<+label+>}
\end{align}
Thus, the Hamiltonian density, found from the Lagrangian using a Legendre transformation, reads
\begin{equation}
  \mathcal{H}
  = \Pi (\partial_0 \Psi) - \Pi^* (\partial_0 \Psi^*) - \mathcal{L}
  = \frac{\hbar^2}{2 m} \partial_i (\Psi^{*} \partial^i \Psi) - \frac{\hbar^2}{2 m} \Psi^{*} \partial_i \partial^i \Psi + \Psi^{*} V \Psi.
  \label{eqn:ordinary_hamiltonian}
\end{equation}
We then observe that, when the Hamiltonian density is integrated over the entire space, thus obtaining the Hamiltonian, we find the expectation value of Eq. \eqref{eqn:ordinary_schrodinger} under the assumption that the fields vanish at infinity.

\subsection{Symmetries: time shifts}

We now proceed by studying the symmetries of the theory, starting with time shifts.
Specifically, let us consider a transformation law for the time coordinate of the form
\begin{equation}
  t \to t - \varepsilon,
  \label{<+label+>}
\end{equation}
with $\varepsilon =$ const.
We can then introduce a conserved current \cite{Dickbook}
\begin{equation}
  j^{i}
  = \varepsilon \left[ - \partial_0 \Psi \frac{\partial \mathcal{L}}{\partial (\partial_i \Psi)} - \partial_0 \Psi^* \frac{\partial \mathcal{L}}{\partial (\partial_i \Psi^*)} \right]
  = \varepsilon \frac{\hbar^2}{2 m} \left(\partial_0 \Psi \partial^i \Psi^{*} + \partial_0 \Psi^* \partial^i \Psi\right),
\end{equation}
and conserved charge
\begin{equation}
  \rho
  = \varepsilon \left[ \mathcal{L} - \partial_0 \Psi \frac{\partial \mathcal{L}}{\partial (\partial_0 \Psi)} - \partial_0 \Psi^* \frac{\partial \mathcal{L}}{\partial (\partial_0 \Psi^*)} \right]
  = - \varepsilon \mathcal{H}.
\end{equation}
We thus conclude that the conserved quantity in a time shift is the Hamiltonian (density).

\subsection{Symmetries: spatial translations}

Let us repeat the same argument for the case of spatial translations, that is, for a transformation of coordinates of the form
\begin{equation}
  x^i \to x^i - \varepsilon^i,
  \label{eqn:spatial_translation}
\end{equation}
with $\varepsilon^i =$ const for $i = 1,2,3$.
Thus, we can identify a conserved current \cite{Dickbook}
\begin{multline}
  j^{i}
  = \varepsilon^k \left[ \eta_k^i \mathcal{L} - \partial_k \Psi \frac{\partial \mathcal{L}}{\partial (\partial_i \Psi)} - \partial_k \Psi^* \frac{\partial \mathcal{L}}{\partial (\partial_i \Psi^*)} \right]\\
  = \varepsilon^i \left[ \frac{i \hbar}{2}\left(\Psi^{*} \partial_0 \Psi - \partial_0 \Psi^{*} \Psi\right) - \Psi^{*} V \Psi \right]
  + \varepsilon^k \frac{\hbar^2}{2 m} \left(\partial^i \Psi^* \partial_k \Psi + \partial_k \Psi^* \partial^i \Psi\right)
\end{multline}
and conserved charge
\begin{multline}
  \rho
  = \varepsilon^k \left[ - \partial_k \Psi \frac{\partial \mathcal{L}}{\partial (\partial_0 \Psi)} - \partial_k \Psi^* \frac{\partial \mathcal{L}}{\partial (\partial_0 \Psi^*)} \right]
  = - \varepsilon^k \frac{i \hbar}{2} \left[ \Psi^* \partial_k \Psi - \Psi \partial_k \Psi^* \right]\\
  = - \varepsilon^k i \hbar \Psi^* \partial_k \Psi + \frac{i \hbar}{2} \partial_k \left(\varepsilon^k |\Psi|^2\right).
\end{multline}
The last term, when integrated over the entire space, cancels out under the assumption of vanishing fields at infinity.
Therefore, we can conclude that the conserved quantities under spatial translations are the components of the momentum.

\subsection{Symmetries: complex phase shifts}

Finally, let us consider a transformation which changes the fields by a complex phase $\varphi$, that is
\begin{equation}
  \Psi \to \Psi' 
  = e^{i \varphi} \Psi
  \simeq \Psi (1 + i \varphi).
  \label{eqn:phase_shift}
\end{equation}
In this case, the current density is \cite{Dickbook}
\begin{equation}
  j^i
  = - \delta \Psi \frac{\partial \mathcal{L}}{\partial (\partial_i \Psi)} - \delta \Psi^* \frac{\partial \mathcal{L}}{\partial (\partial_i \Psi^*)}
  = - i \varphi \frac{\hbar^2}{2m} \left(\Psi^* \partial^i \Psi - \Psi \partial^i \Psi^*\right),
\end{equation}
while for the conserved charge we get
\begin{equation}
  \rho
  = - \delta \Psi \frac{\partial \mathcal{L}}{\partial (\partial_0 \Psi)} - \delta \Psi^* \frac{\partial \mathcal{L}}{\partial (\partial_0 \Psi^*)}
  = \varphi \hbar |\Psi|^2.
\end{equation}
Therefore, these are the probability current density and probability density, respectively, for a field (wavefunction) $\Psi$.

\section{Models of Quantum Mechanics with a minimal uncertainty in position}
\label{sec:MLQM_review}

Here, we will summarize the main features of models of minimal-length quantum mechanics.
Thus, let us suppose that the minimal length is obtained via a modified commutator of the form
\begin{equation}
  [\hat{x}^{a},\hat{p}_{b}] 
  = i \hbar \left[f(\hat{p}^{2}) \delta_{b}^{a}+\bar{f}(\hat{p}^{2})\frac{\hat{p}^{a}\hat{p}_{b}}{\hat{p}^{2}}\right]
  = i \hbar F^a_b(\vec{\hat{p}})
  \label{eqn:GUP_commutator}
\end{equation}
with suitable choices of the functions $f$ and $\bar{f}$ \cite{Bosso:2023aht}.
In particular, demanding that the coordinates commute, as we will do in the following, we require
\begin{equation}
  \bar{f}
  = \frac{2 f f' \hat{p}^2}{f - 2 f' \hat{p}^2}.
  \label{<+label+>}
\end{equation}
Within such models, the Hamiltonian can be written as \cite{Bosso:2023nst}
\begin{equation}
  \hat{H} = \frac{\hat{p}^2}{2 m} + V(|\vec{\hat{X}}|),
  \label{eqn:GUP_Hamiltonian}
\end{equation}
where $X^a$ is the conjugate configuration variable to $p_b$, namely \cite{Bosso:2018uus}
\begin{equation}
  [\hat{X}^a,\hat{p}_b] = i \hbar \delta^a_b,
  \label{eqn:Xp_commutator}
\end{equation}
and $X^a$ approximates to the ordinary position $x^a$ in the low-energy limit.
It is worth pointing out that the momentum $\vec{p}$ has a linear composition law.
Furthermore, it is possible introducing the wavenumber operator $\vec{\hat{k}}$ as the conjugate momentum to the position $\vec{\hat{x}}$, that is, such that \cite{Bosso:2020aqm,Bosso:2023sxr}
\begin{align}
  \hat{k}^a
  &= \frac{1}{\hbar} \bar{g}(\hat{p}^2) \hat{p}^a, &
  [\hat{x}^a,\hat{k}_b]
  &= i \delta^a_b.
  \label{eqn:wavenumber_momentum}
\end{align}
For a minimal length to be present, the wavenumber operator has to be bounded \cite{Bosso:2023sxr}.
Consequently, and in contrast to $\vec{p}$, the wavenumber $\vec{k}$ cannot compose linearly.
We will therefore call $\kappa$ the set of its eigenvalues.

In what follows, it is also useful considering the operator $\hat{p}_a$ as a function of the vector operator $\vec{\hat{k}}$.
In particular, inverting Eq. \eqref{eqn:wavenumber_momentum}, we can write
\begin{equation}
  \hat{p}_a = \hbar g(\hat{k}^2) \hat{k}_a,
  \label{eqn:momentum_wavenumber}
\end{equation}
with $g(k^2) = 1 / \bar{g}(p^2)$.
From the assumption of commutativity, it is possible to relate the functions $\bar{g}$ and $f$ \cite{Bosso:2023sxr}.
In particular, we have
\begin{equation}
  \bar{g}(p^2)
  = \frac{1}{f(p^2)}\qquad
  g(k^2) = f(p^2).
  \label{<+label+>}
\end{equation}

To find the relation between the coordinates $X^a$ and $x^a$, from Eq.\eqref{eqn:GUP_commutator} and noting that $F^a_b$ is a function of $\vec{p}$ only, we have
\begin{equation}
  x^b
  = \frac{1}{2} \left\{F^b_a(\vec{\hat{p}}), \hat{X}^a\right\},
  \label{eqn:x_X}
\end{equation}
where $\{\cdot,\cdot\}$ indicates an anticommutator.
Furthermore, from Appendix A of \cite{Bosso:2021koi}, it is easy to show that
\begin{equation}
  [\hat{X}^a,\hat{k}_b]
  = (F^{-1})^a_b(\vec{\hat{p}})
  = (G^{-1})^a_b(\vec{\hat{k}})
  \label{<+label+>}
\end{equation}
where we introduced the matrix operator \mbox{$G^a_b(\vec{\hat{k}}) = F^a_b(\vec{\hat{p}})$}.
From this, we can write
\begin{equation}
  \hat{X}^a
  = \frac{1}{2} \left\{(G^{-1})^a_b(\vec{\hat{k}}), \hat{x}^b\right\}.
  \label{eqn:X_x}
\end{equation}
It is worth mentioning that the symmetrical ordering in Eqs. \eqref{eqn:x_X} and \eqref{eqn:X_x} allows for the ordinary measure for momentum space, although other ordering prescriptions can be equivalently applied \cite{Bosso:2021koi}.
Furthermore, it is useful finding the determinant of the matrices $F^a_b$ and $G^a_b$.
We have, respectively,
\begin{align}
  |F|
  &= \frac{f^4}{f - 2 f' p^2},&
  |G|
  &= g^2 (g + 2 g' k^2).
  \label{<+label+>}
\end{align}
Using these expressions explicitly, we find the following useful relations
\begin{align}
  \frac{\partial}{\partial k_b} |G| (G^{-1})^a_b &= 0 &
  \frac{\partial}{\partial p_b} \frac{F^a_b}{|F|} &= 0.
  \label{<+label+>}
\end{align}

In what follows, we will often describe a system in the representation in which either the operators $\hat{X}^a$ or $\hat{x}^a$ are multiplicative operators.
The transformation rules between such spaces and the representations for the various operators can be found in Appendix \ref{apx:representations}.

\subsection{Example: Kempf-Mangano-Mann model}
\label{sec:KMM}

\begin{figure}
  \centering
  \includegraphics[width=\columnwidth]{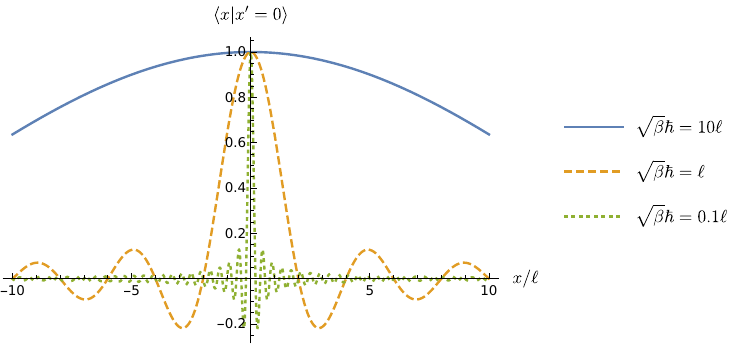}
  \caption{Inner product between two eigenstates of the operator $x$.
  The parameter $\beta$ has been chosen so to produce a minimal length equal to $10 \ell$, $\ell$, and $0.1 \ell$, where $\ell$ is an arbitrary length scale.}
  \label{fig:x_x}
\end{figure}

As a toy model, we will consider the one-dimensional deformation considered in \cite{Kempf:1994su}, namely
\begin{equation}
  [\hat{x}, \hat{p}]
  = i \hbar (1 + \beta \hat{p}^2).
  \label{eqn:KMM}
\end{equation}
Such a model is characterized by a bounded wavenumber operator according to
\begin{equation}
  |k| < \frac{\pi}{2 \sqrt{\beta} \hbar}
  \label{<+label+>}
\end{equation}
and therefore a minimal uncertainty in position given by \cite{Bosso:2023sxr}
\begin{equation}
  \Delta x \geq \sqrt{\beta} \hbar.
  \label{<+label+>}
\end{equation}
In this model, we have
\begin{align}
  g(\hat{k}^2) &= \frac{\tan(\sqrt{\beta} \hbar \hat{k})}{\sqrt{\beta} \hbar \hat{k}},&
  \bar{g}(\hat{p}^2) &= \frac{\arctan(\sqrt{\beta} \hat{p})}{\sqrt{\beta} \hat{p}}.
  \label{eqn:g_barg}
\end{align}
Furthermore, in $p$-space representation, the coordinate operators, in symmetrical ordering, are
\begin{align}
  \hat{x}
  &= i \hbar (1 + \beta p^2) \frac{\diff}{\diff p} + i \hbar \beta p,&
  \hat{X} &= i \hbar \frac{\diff}{\diff p}.
  \label{<+label+>}
\end{align}
It is possible to use this relation to find the $x$-eigenfunctions and the corresponding inner product.
We thus find
\begin{align}
  \langle p | x \rangle
  &= \sqrt{\frac{\sqrt{\beta }}{\pi}} \frac{e^{-\frac{i x \arctan\left(\sqrt{\beta } p\right)}{\sqrt{\beta } \hbar }}}{\sqrt{1 + \beta  p^2}},\\
  \langle x | x' \rangle
  &= \frac{2 \sqrt{\beta} \hbar \sin \left[\frac{\pi (x-x')}{2 \sqrt{\beta } \hbar }\right]}{\pi (x- x')}.
  \label{<+label+>}
\end{align}
The inner product is presented in Fig. \ref{fig:x_x}.
It is worth observing that the $x$-eigenstates are not orthogonal.

\section{Minimal length classical field theory}
\label{sec:MLCFT}

Now, we are going to describe the model introduced in the previous section in terms of a classical field theory.
In particular, remembering that $p_a$ and $X^a$ form a conjugate pair and considering that we expect a Schr\"odinger-like equations in these two quantities, as hinted by the Hamiltonian in Eq. \eqref{eqn:GUP_Hamiltonian}, the Lagrangian corresponding to this model, by comparison with the ordinary case in Eq. \eqref{eqn:ordinary_lagrangian}, is
\begin{equation}
  \mathcal{L}
  = \frac{i \hbar}{2}\left(\Psi^{*} \partial_0 \Psi - \partial_0 \Psi^{*} \Psi\right) - \frac{\hbar^2}{2 m} \dot{\partial}_i \Psi^{*} \dot{\partial}^i \Psi-\Psi^{*} V \Psi,
  \label{eqn:GUP_Lagrangian}
\end{equation}
where $\dot{\partial_i} = \frac{\partial}{\partial X^i}$.
Although it has the same form as the ordinary Lagrangian, and therefore admits the same equations of motion in terms of the variables $\{X^a,p^a\}$, the description in terms of the ordinary position is far from being straightforward since the functional relation between the two sets of coordinates involves the momentum $\vec{p}$ or, alternatively, the wavenumber $\vec{k}$ \cite{Bosso:2022vlz}.
Thus, changing from $X^a$ to $x^a$ results in a higher (potentially infinite, as in the case of the model in Section \ref{sec:KMM}) order Lagrangian.
Nonetheless, employing the similarity between the Lagrangians in Eqs. \eqref{eqn:ordinary_lagrangian} and \eqref{eqn:GUP_Lagrangian}, we have that the conserved quantity in a time shift is the Hamiltonian, which in the present context, up to a total derivative, acquires the same form as in Eq. \eqref{eqn:ordinary_hamiltonian}
\begin{equation}
  \mathcal{H}
  = - \frac{\hbar^2}{2 m} \Psi^{*} \dot{\partial}_i \dot{\partial}^i \Psi + \Psi^{*} V(|\vec{X}|) \Psi.
  \label{eqn:GUP_Hamiltonian_X}
\end{equation}
or, in $x$-space,
\begin{equation}
  \mathcal{H}
  = - \frac{\hbar^2}{2m} \Psi^* g^2(-\nabla^2) \partial_i \partial^i \Psi 
  + \Psi^{*} V\left( \frac{1}{2} \left\{ (G^{-1})^i_a (-i\vec{\nabla}), x^a \right\} \right) \Psi. 
  \label{eqn:GUP_Hamiltonian_x}
\end{equation}
Here, the symbol $g(-\nabla^2)$ is understood as a (infinite) series of derivative, that is,
\begin{equation}
  g(-\nabla^2)
  = \sum_{j=0}^{\infty} \frac{(-1)^j}{j!} \left[ \frac{\diff^j g(k^2)}{\diff (k^2)^j} \right]_{k^2=0} (\nabla^2)^{j}.
  \label{<+label+>}
\end{equation}
The Hamiltonian can be found by introducing the momenta
\begin{align}
  \Pi
  &= \frac{\diff \mathcal{L}}{\diff (\partial_0 \Psi)}, &
  \Pi^* 
  &= \frac{\diff \mathcal{L}}{\diff (\partial_0 \Psi^*)},
  \label{eqn:field_momenta}
\end{align}
which have the same identical form as in the ordinary theory.
It is worth observing that, to derive the above Hamiltonian, we do not need to introduce any other field, nor we need to modify the Poisson brackets between the fields and their conjugate momenta.
The essence of the minimal length, which can be traced back to Eqs. \eqref{eqn:wavenumber_momentum} and \eqref{eqn:momentum_wavenumber}, is encapsulated in the specific form of the Lagrangian or the corresponding Hamiltonian in the two representations, namely Eqs. \eqref{eqn:GUP_Hamiltonian_X} and \eqref{eqn:GUP_Hamiltonian_x}.

Although the description of time shifts trivially descends from that in the ordinary theory, that of spatial translations is not this straightforward.
Indeed, consider a transformation of the form \eqref{eqn:spatial_translation}.
As a canonical transformation, this leaves the momentum $k^a$, and therefore $p^a$, as is.
However, when described in terms of the variables $X^a$, it corresponds to a transformation involving both $X^a$ and $p^a$.

To have a better insight, let us start by considering a free Lagrangian, which in terms of the position $x^a$ and derivatives $\partial_a$ reads
\begin{equation}
  \mathcal{L}
  = \frac{i \hbar}{2}\left(\Psi^{*} \partial_0 \Psi - \partial_0 \Psi^{*} \Psi\right) 
  - \frac{\hbar^2}{2 m} [g(-\nabla^2) \partial_a \Psi^{*}] [g(-\nabla^2) \partial^a \Psi],
  \label{eqn:GUP_Lagrangian_x}
\end{equation}
Thus, in terms of $x^a$, the Lagrangian presents higher derivative terms.

Using the stationary action principle, we can then write
\begin{equation}
  \frac{\delta S}{\delta \Psi^*}
  = \sum_{j=0}^{\infty} (-1)^j \partial_{\mu_1 \ldots \mu_j} \frac{\partial \mathcal{L}}{\partial \Psi_{\mu_1 \ldots \mu_j}^*}
  = 0,
  \label{eqn:eom_Psi}
\end{equation}
in which, using standard notation, we have written
\begin{align}
  \partial_{\mu_1 \ldots \mu_j} &= \partial_{\mu_1} \ldots \partial_{\mu_j},&
  \Psi_{\mu_1 \ldots \mu_j}
  &= \partial_{\mu_1 \ldots \mu_j} \Psi.
  \label{<+label+>}
\end{align}
We can now check whether the current description is indeed equivalent to the one given by the free counterpart of Eq. \eqref{eqn:GUP_Lagrangian}.
For example, the equation of motion becomes
\begin{equation}
  0
  = i \hbar \partial_0 \Psi
  + \frac{\hbar^2}{2 m} g^2(-\nabla^2) \partial_a \partial^a \Psi.
  \label{eqn:eom_X}
\end{equation}
Notice that, in terms of $X^a$ and $\dot{\partial}_a$, the equations of motion is
\begin{equation}
  0
  = i \hbar \partial_0 \Psi + \frac{\hbar^2}{2m} \dot{\partial}_a \dot{\partial}^a \Psi.
  \label{eqn:eom_x}
\end{equation}
As expected, the equations of motion descried in terms of $x^a$ and $X^a$ are equivalent when we take Eq. \eqref{eqn:momentum_wavenumber} in consideration.
Furthermore, they correspond to the Schr\"odinger equation in the two representations.
Finally, it is easy to see that the same equivalence between the two formulations is maintained even when a potential $V(\vec{X})$ is present.

\section{Symmetries}
\label{sec:symmetries}

We will dedicate this section to the study of symmetry transformations in the context of minimal length models.
We will focus on spatial translations and phase shifts, as time shifts have been already considered in the previous section.
For this purpose, a generic expression for the conserved current and density associated to an arbitrary transformation are derived in Appendix \ref{apx:current_MLQM} for the case of spatial derivatives of any order.

\subsection{Spatial translations}
\label{sec:MLCFT_spatial_translations}

Let us thus consider a transformation of the form of Eq. \eqref{eqn:spatial_translation}.
By using Eq. \eqref{eqn:GUP_current}, or alternatively \eqref{eqn:energy_momentum}, and the Lagrangian in Eq. \eqref{eqn:GUP_Lagrangian_x}, we find for the conserved charge
\begin{equation}
  \pi_i
  = \Theta_i{}^0
  = - \frac{\partial \mathcal{L}}{\partial \Psi_{0}} \partial_i \Psi - \frac{\partial \mathcal{L}}{\partial \Psi^*_{0}} \partial_i \Psi^*
  = - \frac{i \hbar}{2} \left(\Psi^* \partial_i \Psi  - \Psi \partial_i \Psi^*\right).
  \label{eqn:conserved_momentum}
\end{equation}
Thus, as expected, the conserved charge under spatial translation is the wavenumber $\vec{k}$ and is consistent with ordinary quantum mechanics.

\subsection{Complex phase shifts}
\label{sec:MLCFT_phase_shifts}

Let us now consider a transformation of the form given in Eq. \eqref{eqn:phase_shift}.
For the conserved current, we then have
  \begin{multline}
    J^{i}
    = - \frac{i \hbar}{2 m} \sum_{N=0}^{\infty} \sum_{n=0}^{N} \sum_{l=1}^{2n+1} \sum_{\substack{n_1,n_2,n_3=0\\n_1+n_2+n_3=n}}^{n} \frac{(-1)^{l+N}}{n_1! n_2! n_3! (N-n)!}\\
    \times \left[ \frac{\diff^{n} g(k^2)}{\diff (k^2)^{n}} \right]_{k^2=0} \left[ \frac{\diff^{N-n} g(k^2)}{\diff (k^2)^{N-n}} \right]_{k^2=0} \xi_{a \underbrace{\scriptstyle1\ldots1}_{2n_1\text{times}} \underbrace{\scriptstyle2\ldots2}_{2n_2\text{times}} \underbrace{\scriptstyle3\ldots3}_{2n_3\text{times}}}^{i\mu_2\ldots\mu_{2n+1}}\\
    \times \left[ \left( \partial_{\mu_2\ldots\mu_{l}} \nabla^{2(N-n)} \partial^i \Psi^{*} \right) \left( \partial_{\mu_{l+1}\ldots\mu_{2n+1}} \Psi \right) \right.\\
    \left. - \left( \partial_{\mu_2\ldots\mu_{l}} \nabla^{2(N-n)} \partial^a \Psi \right) \left( \partial_{\mu_{l+1}\ldots\mu_{2n+1}} \Psi^* \right) \right]
    \label{eqn:GUP_probability_current}
  \end{multline}
up to a factor $\hbar \varphi$ associated with the arbitrary phase shift and where we introduced the symbol
\begin{equation}
  \xi^{a_1 \ldots a_n}_{b_1 \ldots b_n} = \begin{cases}
    1 & \text{when } \{a_i\}_{i=1}^n = \{b_i\}_{i=1}^n;\\
    0 & \text{otherwise,}
  \end{cases}
  \label{<+label+>}
\end{equation}
which is completely symmetric with respect to any arbitrary exchange of the indices in the two sets.
Besides the complicated-looking form, it is worth observing that, from a perturbative perspective, the integer $N$ in Eq. \eqref{eqn:GUP_probability_current} corresponds to the expansion order of the probability current, as we will shortly see.
In particular, for $N=0$ we have
\begin{equation}
  J^{i} 
  = \frac{i \hbar}{2 m} \left[ \left( \partial^i \Psi^{*} \right) \Psi - \left( \partial^i \Psi \right) \Psi^* \right]
\end{equation}
which corresponds to the probability current density of ordinary quantum mechanics.

As for the probability density, we find, to all orders,
\begin{equation}
  \rho 
  = - \frac{i}{\hbar} \left(\frac{\partial \mathcal{L}}{\partial \Psi_{0}} \Psi - \frac{\partial \mathcal{L}}{\partial \Psi_{0}} \Psi^*\right)
  = |\Psi|^2,
  \label{eqn:probability_density}
\end{equation}
corresponding to the usual probability density.

\subsection{Field theory example: Kempf-Mangano-Mann model}

We will now apply the results presented above to the model introduced in Sec. \ref{sec:KMM}.
We start with the free, one-dimensional Lagrangian
\begin{equation}
  \mathcal{L}
  = \frac{i \hbar}{2} \left(\Psi^{*} \partial_0 \Psi - \partial_0 \Psi^{*} \Psi\right) 
  - \frac{\hbar^2}{2 m} g(- \partial^2) \partial \Psi^{*} g(- \partial^2) \partial \Psi.
\end{equation}
In this case, as shown in Appendix \ref{apx:1D_current}, the probability current is
\begin{equation}
  J 
  = \frac{i \hbar}{2 m} \sum_{N=0}^{\infty} \frac{(-1)^N}{N!} \left[ \frac{\diff^{N}}{\diff x^{N}} g^2(k^2) \right]_{x=0}
  \sum_{l=0}^{2N+1} (-1)^{l} \left( \partial^{(2N-l+1)} \Psi^{*} \right) \left( \partial^{(l)} \Psi \right).
\end{equation}

Inserting explicitly the expressions presented in Sec. \ref{sec:KMM} and expanding up to first order in $\beta$ ($N=1$), we find
\begin{multline}
  J 
  = \frac{i \hbar}{2 m} \left[ \left( \partial \Psi^{*} \right) \Psi - \left( \partial \Psi \right) \Psi^* \right]\\
  - \frac{i \hbar}{3 m} \beta \hbar^2 \left[ \left( \partial^{(3)} \Psi^{*} \right) \Psi - \left( \partial^{(2)} \Psi^* \right) \left( \partial \Psi \right)
  + \left( \partial \Psi^{*} \right) \left( \partial^{(2)} \Psi \right) - \Psi^* \left( \partial^{(3)} \Psi \right) \right].
\end{multline}
This expression is consistent with what was previously found in \cite{Das:2009hs}.

\section{Second quantization}
\label{sec:MLQFT}

Above, we have seen that ordinary and minimal-length quantum mechanics are equivalent to a classical field theory described in terms of the field $\Psi$, its canonical momentum $\Pi$, and the corresponding complex conjugate fields, $\Psi^*$ and $\Pi^*$.
We now proceed to second-quantize such fields, imposing a commutation relation, focusing on a free system, as interactions can easily be introduced following the arguments included in Sec. \ref{sec:MLCFT}.

As pointed out above, the fields $\Psi$ and $\Psi^*$ form a pair of canonically conjugate fields, while the information on a minimal measurable length is already present in the specific form of the Lagrangian and the bounded character of the wavenumber $\vec{k}$.
In other words, there is no need to modify the Poisson brackets between the fields.
As a consequence, we are going to impose ordinary commutation relations for the fields, resulting in
\begin{align}
  [\Psi(\vec{X},t),\Psi^*(\vec{X}',t')]
  = \delta^{(3)}(\vec{X} - \vec{X}') \delta(t - t'),
  \label{eqn:commutation_fields}
  \\[1em]
  [\Psi(\vec{X},t), \Psi(\vec{X}',t')]
  = [\Psi^*(\vec{X},t), \Psi^*(\vec{X}',t')]
  = 0.
  \label{<+label+>}
\end{align}
It is worth observing that, given the momenta in Eq. \eqref{eqn:field_momenta} and the Lagrangian in Eq. \eqref{eqn:GUP_Lagrangian}, the relations between fields and the corresponding momenta are not modified, that is
\begin{equation}
  [\Psi(\vec{X},t),\Pi^*(\vec{X}',t')]
  = \frac{i \hbar}{2} \delta^{(3)}(\vec{X} - \vec{X}') \delta(t - t').
  \label{<+label+>}
\end{equation}
In terms of the coordinates $x^a$, using Eq. \eqref{eqn:x_X}, we have
\begin{equation}
  [\Psi_x(\vec{x},t),\Psi_x^*(\vec{x}',t')]
  = \langle \vec{x} | \vec{x}' \rangle \delta(t - t'),
\end{equation}
where, to not clutter the expression, we have used the notation $\langle \vec{x}|\vec{x}'\rangle$ to represent the function corresponding to the scalar product between $\vec{x}$-eigenstates at $\vec{x}$ and $\vec{x}'$, presented in Eq. \eqref{eqn:inner_product_general}.
We then see that the commutation relations between fields is modified when described in terms of the coordinates $x^a$.
However, since $\langle \vec{x} | \vec{x}' \rangle \to \delta(\vec{x} - \vec{x}')$ in the limit of ordinary quantum mechanics, we recover the ordinary quantum field theory description in the limit of vanishing minimal position uncertainty.

As typical in both quantum field theory and in models of minimal length, a description in momentum space is convenient.
The question to ask at this point concerns the choice of pairs of canonical variables to be used, that is, if one should consider a Fourier transform between $\vec{X}$ and $\vec{p}$ or between $\vec{x}$ and $\vec{k}$.
We claim that the first is the route to pursue since, as hinted in Sec. \ref{sec:MLQM_review} and in \cite{Bosso:2023nst}, the momentum $\vec{p}$ is defined to compose linearly.

Let us then introduce the Fourier amplitudes $a(\vec{p},t)$ and $a^*(\vec{p},t)$ for the classical fields as
\begin{equation}
  a(\vec{p},t)
  = (2 \pi \hbar)^{-3/2} \int \diff^3 X ~ \Psi(\vec{X},t) e^{- i \vec{p} \cdot \vec{X}/\hbar}.
  \label{<+label+>}
\end{equation}
Since the commutation relation in Eq. \eqref{eqn:commutation_fields} is identical to the ordinary one, upon promoting the field modes to operators, we obtain the usual algebra, namely
\begin{align}
  [a(\vec{p},t), a^\dagger (\vec{p}',t)] &= \delta^{(3)}(\vec{p}-\vec{p}'),
  \\[1em]
  [a(\vec{p},t), a(\vec{p}',t')]
  &= [a^\dagger (\vec{p},t), a^\dagger (\vec{p}',t')]
  = 0.
  \label{<+label+>}
\end{align}
For the same reason, we find the following expression for the number operator
\begin{equation}
  \hat{N}
  = \int_{\mathbb{R}^3} \diff^3 X ~ \rho
  = \int_{\mathbb{R}^3} \diff^3 p ~ a^\dagger (\vec{p},t) a(\vec{p},t).
  \label{<+label+>}
\end{equation}
The relation above, as usual in quantum field theory, can be interpret as counting the particles in mode $\vec{p}$ and summing over all modes.
A similar description can be given in terms of the wavenumber $\vec{k}$, employing the relation given in Eq. \eqref{eqn:momentum_wavenumber}.
Specifically, we have
\begin{equation}
  \hat{N}
  = \int_{\kappa} \diff^3 k ~ \hbar^3 |G| a_k^\dagger(\vec{k},t) a_k(\vec{k},t),
  \label{eqn:number_operator_k}
\end{equation}
with 
\begin{equation}
  a_k(\vec{k},t)
  = a(\vec{p},t).
  \label{<+label+>}
\end{equation}
The operators $a_k$ and $a^\dagger_k$ are good ladder operators in $k$-space.
Indeed, we find
\begin{equation}
  [a_k(\vec{k},t), a_k^\dagger (\vec{k}',t)]
  = [a(\vec{p},t), a^\dagger (\vec{p}',t)]
  = \delta(\vec{p}-\vec{p}')
  = \frac{1}{\hbar^3 |G|} \delta(\vec{k}-\vec{k}'),
\end{equation}
which agrees with Eq. \eqref{eqn:inner_product_k}.
In this sense, the number operator written in terms of the ladder operators in $k$-space can be interpreted as usual as counting the number of particle in mode $\vec{k}$ and then summing over all modes, with the further care of correcting the count by the measure factor $\hbar^3 |G|$.

As for the momentum operator, we can write
\begin{equation}
  \hat{p}_a
  = - i \hbar \int_{\mathbb{R}^3} \diff^3 X ~ \Psi^\dagger(\vec{X},t) \dot{\partial}_a \Psi(\vec{X},t)
  = \int_{\mathbb{R}^3} \diff^3 p ~ a^\dagger(\vec{p},t) a(\vec{p},t) p_{a},
\end{equation}
which follows the ordinary interpretation of counting particles and summing over all modes.
As for the wavenumber $\vec{k}$, following Eq. \eqref{eqn:wavenumber_momentum}, we defined the corresponding operator as
\begin{equation}
  \hat{k}_a
  = \frac{1}{\hbar} \bar{g}(\hat{p}^2) \hat{p}_a.
  \label{<+label+>}
\end{equation}
It is worth noticing that, in this case, the usual ``count and sum'' interpretation cannot be adopted as $\vec{k}$ does not compose linearly.

\subsection{Example: quartic interaction}
\label{sec:MLQFT_quartic}

As an example, let us consider a scalar system characterized by a quartic interaction
\begin{equation}
  \mathcal{L}
  = \Psi^\dagger \left(i \hbar \partial_0 + \frac{\hbar^2}{2 m} \dot{\partial}_i \dot{\partial}^i\right) \Psi - \frac{\lambda}{2} \Psi^\dagger \Psi^\dagger \Psi \Psi.
\end{equation}
Following the standard procedure for this problem, the Hamiltonian can be written in Fourier components as the sum of a free term $H_0$ and an interaction term $H_{\text{int}}$
\begin{equation}
  H_0
  = \frac{1}{2 m} \int \diff^3 p ~ a^\dagger(\vec{p},t) a(\vec{p}',t)  p_i p^i,
\end{equation}
\begin{multline}
  H_{\text{int}}
  = \frac{\lambda}{16(\pi \hbar)^3} \int \diff p_1^3 \diff p_2^3 \diff p_3^3 \diff p_4^3 ~ a^\dagger(\vec{p}_4,t) a^\dagger(\vec{p}_3,t) \\
   \qquad \times a(\vec{p}_2,t) a(\vec{p}_1,t) \delta^3(\vec{p}_1 + \vec{p}_2 - \vec{p}_3 - \vec{p}_4).
  \label{<+label+>}
\end{multline}
At this point, one is usually interested in the matrix elements of $H_{\text{int}}$.
Let us then consider the interaction between two particles described initially by a state of definite momenta $|\vec{p}'_1,\vec{p}'_2\rangle$ and evolving into another state of definite momenta $|\vec{p}'_3,\vec{p}'_4\rangle$.
The matrix elements for the interaction are then
\begin{equation}
  \langle \vec{p}'_3, \vec{p}'_4 | H_{\text{int}} | \vec{p}'_1, \vec{p}'_2\rangle
  = \frac{\lambda}{4(\pi \hbar)^3} \delta^{(3)}[(\vec{p}_1' + \vec{p}_2') - (\vec{p}_3' + \vec{p}_4')].
\end{equation}
As expected, we get the same result as in the standard theory and, in particular, momentum is conserved in the interaction.

\begin{figure}
  \centering
  \includegraphics[width=\columnwidth]{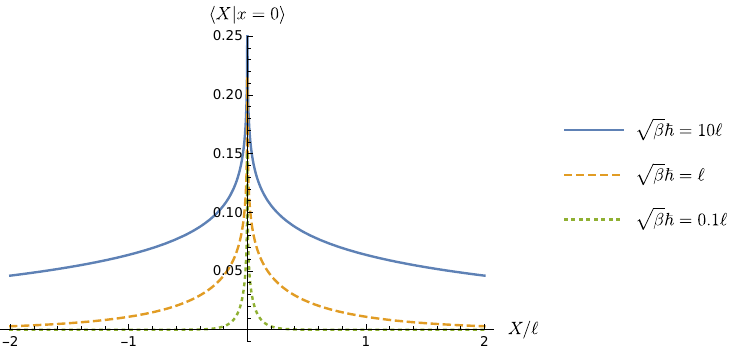}
  \caption{Function $\langle X | x = 0\rangle$ for the model of Sec. \ref{sec:KMM}.
    Four different values of $\beta$ have been considered, corresponding to values of the minimal length equal to $10 \ell_{\text{P}}$, $\ell$, $0.1 \ell$, with $\ell$ an arbitrary length scale.
  For presentation reasons, the functions are normalized so that $\langle X=0 | x=0 \rangle = 1$.}
  \label{fig:X_x}
\end{figure}

Let us now consider the following matrix elements, corresponding to the interaction between two particles initially at $\vec{X}_1$ and $\vec{X}_2$ and then ending up in $\vec{X}_3$ and $\vec{X}_4$
\begin{multline}
  \langle \vec{X}_3, \vec{X}_4 | H_{\text{int}} | \vec{X}_1, \vec{X}_2\rangle\\
  = 2 \lambda \int \diff^3 X ~ \delta^{(3)}(\vec{X} - \vec{X}_4) \delta^{(3)}(\vec{X} - \vec{X}_3) 
  \delta^{(3)}(\vec{X} - \vec{X}_2) \delta^{(3)}(\vec{X} - \vec{X}_1).
\end{multline}
This result simply states that the interaction is local in the $\vec{X}$-space, that is, the interaction is relevant only when $\vec{X}_1 = \vec{X}_2 = \vec{X}_3 = \vec{X}_4$, as in the ordinary theory.
However, we stress that $\vec{X}$ does not represent a position in ordinary space.
To know what happens for the ordinary position $\vec{x}$, we compute the following
\begin{equation}
  \langle \vec{x}_3, \vec{x}_4 | H_{\text{int}} | \vec{x}_1, \vec{x}_2\rangle
  = 2 \lambda \int \diff^3 X ~ \langle \vec{x}_3 | \vec{X} \rangle \langle \vec{x}_4 | \vec{X} \rangle \langle \vec{X} | \vec{x}_2 \rangle \langle \vec{X} | \vec{x}_1 \rangle.
\end{equation}
Now, it is easy to see that the function $\langle \vec{X} | \vec{x} \rangle$, considered as a function of $\vec{X}$ for any specific choice of $\vec{x}$, is a wider function than a Dirac delta, as demonstrated in Fig. \ref{fig:X_x} for the specific model of Sec. \ref{sec:KMM}.
Therefore, the integral above would in general be different from zero every time that the supports of the four functions in the integrand intersect.
This may happens also when $\vec{x}_1 \neq \vec{x}_2 \neq \vec{x}_3 \neq \vec{x}_4$.
Thus, within the framework presented in this work, a quartic interaction is compatible with two particles interacting even when not at the same position, and the outgoing particles can indeed appear in other two different positions.
All this is consistent with a theory characterized by a minimal uncertainty in position.

\section{Conclusions}
\label{sec:conclusions}

Many approaches to quantum gravity present some sort of minimal length.
Phenomenological models including a minimal length have been commonly used to study this feature on low-energy systems.
One of such models consists in modifying the commutation relation between position and momentum to deform the uncertainty relation accordingly.
Although this procedure fits the scope of the program of quantum gravity phenomenology in the context of non-relativistic quantum mechanics, it may be ill-suited to different frameworks, such as quantum field theory.
In this work, we have surpassed this impasse by first resorting to a classical field theory description of quantum mechanics and then proceeding to a second quantization, first-principle introduction of a quantum field theory with a minimal length \emph{de facto}.

Specifically, treating quantum mechanics as a classical field theory, we were first able to write the appropriate Lagrangian and Hamiltonian with a minimal length.
At the same time, this procedure allowed us to identify the correct conserved currents and densities associated with symmetry transformations, and in particular for spatial translations, time shifts, and complex phase shifts.
We then continued by quantizing the model thus far developed introducing a model in non-relativistic quantum field theory compatible with a minimal length.
To show this explicitly, we finally proposed the specific case of a quartic interaction, which in the ordinary theory is associated with an interaction between two particles at a specific point.
However, in this case, it resulted in particles interacting even at different positions, compatibly with the existence of a minimal length.

Using this procedure, it becomes evident that, to obtain a quantum field theory including a minimal length, no modification to the field commutators is required, since the information regarding the presence of a minimal length is embedded in the form of the Lagrangian and in the relations between the variables $\vec{X}$ and $\vec{x}$, the latter being the physical position.
At the same time, this work sets the stage for phenomenological studies related to the nature and implications of a minimal length in quantum field theory, both in non-relativistic, as presented in this paper, and in relativistic contexts, which we hope to treat in future works.

\backmatter

\begin{appendices}

\section{Change of basis}
\label{apx:representations}

Here, we want to describe the change of basis corresponding to the variables $\vec{X}$, $\vec{p}$, $\vec{x}$, and $\vec{k}$.

\subsection{$\vec{X}$ and $\vec{p}$}
Given their definition, it is straightforward to see that the variables $\vec{X}$ and $\vec{p}$ have to be related as in the ordinary theory.
In particular, given Eq.\eqref{eqn:Xp_commutator} and using the ordinary measure for both the $X$- and $p$-spaces, the two bases are related by the ordinary Fourier transform.
Consequently, the operators corresponding to $\vec{X}$ and to $\vec{p}$ in the two representations acquire the usual form.
Furthermore, since $\vec{x}$ and $\vec{k}$ are given by Eqs.\eqref{eqn:x_X} and \eqref{eqn:wavenumber_momentum}, respectively, it is easy to find the corresponding representations in $X$- and $p$-spaces.
Namely, in $X$-space, we have
\begin{align}
  \hat{X}^j &= X \cdot,&
  \hat{p}_j &= - i \hbar \frac{\partial}{\partial X^j},\\
  \hat{x}^j &= \frac{1}{2} \left\{ F^j_a(- i \hbar \vec{\dot{\nabla}}), X^a \right\},&
  \hat{k}_j &= - \frac{i}{f(- \hbar^2 \dot{\nabla}^2)} \frac{\partial}{\partial X^j},
  \label{<+label+>}
\end{align}
with $\dot{\nabla} = \frac{\partial}{\partial X^j} \frac{\partial}{\partial X_j}$, while in $p$-space, we have
\begin{align}
  \hat{X}^j &= i \hbar \frac{\partial}{\partial p_j},&
  \hat{p}_j &= p_j \cdot,\\
  \hat{x}^j &= \frac{1}{2} \left\{ F^j_a(\vec{p}), i \hbar \frac{\partial}{\partial p_a} \right\},&
  \hat{k}_j &= \frac{1}{\hbar f(p^2)} p_j \cdot.
  \label{<+label+>}
\end{align}

\subsection{$\vec{p}$ and $\vec{k}$}

Changing from $\vec{p}$ to $\vec{k}$ or vice versa simply amounts to a change of coordinates.
Thus, besides writing the field in the new coordinates, the integration measure changes according to
\begin{equation}
  \diff^3 p
  \to \diff^3 k \left|\frac{\partial p_b}{\partial k_a}\right|
  = \diff^3 k ~ \hbar^3 |G|.
  \label{<+label+>}
\end{equation}
Thus, the resolution of identity in terms of eigenstates of $\vec{k}$ reads
\begin{equation}
  \mathbb{I}
  = \int_\kappa \diff^3 k ~ \hbar^3 |G| |\vec{k} \rangle \langle \vec{k}|,
  \label{<+label+>}
\end{equation}
while the inner product between such states is
\begin{equation}
  \langle \vec{k} | \vec{k}' \rangle
  = \frac{1}{\hbar^3 |G|} \delta^{(3)}(\vec{k}-\vec{k}').
  \label{eqn:inner_product_k}
\end{equation}
Using these relations and the representations of the operators in $X$- or $p$-spaces, as shown in the previous subsection, we can find the following representation in $k$-space
\begin{align}
  \hat{X}^j &= i (G^{-1})^j_a(\vec{k}) \frac{\partial}{\partial k_a},&
  \hat{p}_j &= \hbar g(k^2) k_j \cdot,\\
  \hat{x}^j &= \frac{1}{2} \left\{ |G|(k^2), \frac{i}{|G|(k^2)} \frac{\partial}{\partial k_b} \right\},\label{eqn:x_in_k} &
  \hat{k}_j &= k_j \cdot.
\end{align}

\subsection{$\vec{k}$ and $\vec{x}$}

To find the correct transform between $k$- and $x$-spaces, we first find the eigenfunctions of the operators $\vec{x}$ in $k$-space, $\langle \vec{k}|\vec{x}\rangle$.
Such functions are solution of the following equation
\begin{equation}
  x^i \langle \vec{k}|\vec{x}\rangle
  = i \frac{\partial}{\partial k_i} \langle \vec{k} | \vec{x} \rangle + \frac{i}{2} \frac{\partial \ln |G|}{ \partial k_i} \langle \vec{k} | \vec{x} \rangle.
  \label{<+label+>}
\end{equation}
We then get
\begin{equation}
  \langle \vec{k} | \vec{x} \rangle
  = \frac{\mathcal{N}_k}{\sqrt{|G|}} e^{-i \vec{x} \cdot \vec{k}}
  \label{<+label+>}
\end{equation}
which $\mathcal{N}_k$ a normalization constant given by
\begin{equation}
  \mathcal{N}_k
  = \frac{1}{\sqrt{V(\kappa) \hbar^3}}
  \label{<+label+>}
\end{equation}
with $V(\kappa) = \int_\kappa \diff^3 k$ the volume of $k$-space.
We can use this relation to find the inner product between $x$-eigenstates.
Specifically,
\begin{equation}
  \langle \vec{x} | \vec{x}' \rangle
  = \frac{1}{V(\kappa)} \int_{\kappa} \diff^3 k ~ e^{i (\vec{x} - \vec{x}') \cdot \vec{k}}.
  \label{eqn:inner_product_general}
\end{equation}

Now, we can find the following relation
\begin{equation}
  \langle \vec{k} | \left[ \int_{\mathbb{R}^3} \diff^3 x ~ |\vec{x}\rangle\langle\vec{x}| \right] |\vec{k}'\rangle
  = \frac{(2 \pi)^3}{V(\kappa)} \langle\vec{k} | \vec{k}'\rangle
  \label{<+label+>}
\end{equation}
from which we have
\begin{equation}
  \int_{\mathbb{R}^3} \diff^3 x ~ |\vec{x}\rangle\langle\vec{x}|
  = \frac{(2 \pi)^3}{V(\kappa)}.
  \label{<+label+>}
\end{equation}
This relation can be used to find the transform from $x$-space to $k$-space, that is, given a state $|\psi\rangle$, we have
\begin{equation}
  \langle \vec{k} | \psi \rangle
  = \frac{\sqrt{V(\kappa)}}{(2 \pi \hbar)^3} \frac{1}{\sqrt{|G|}} \int_{\mathbb{R}^3} \diff^3 x ~ e^{- i \vec{x} \cdot \vec{k}} \langle \vec{x} | \psi \rangle.
\end{equation}
In turn, using this transform we can find the $x$-space representation for the following operators
\begin{align}
  \hat{X}^i
  &= \frac{1}{2} \left\{ (G^{-1})^i_a(-i \vec{\nabla}), x^a \right\},\label{eqn:X_in_x}&
  \hat{p}_i
  &= - i \hbar g(- \nabla^2) \frac{\partial}{\partial x^i},\\
  \hat{x}^i
  &= x \cdot,&
  \hat{k}_i
  &= - i \frac{\partial}{\partial x^i},
  \label{<+label+>}
\end{align}
with $\nabla^2 = \frac{\partial}{\partial x^j} \frac{\partial}{\partial x_j}$.

Finally, the inverse transform is given by
\begin{equation}
  \langle \vec{x} | \psi \rangle
  = \frac{1}{V(\kappa)} \int_{\kappa} \diff^3 k ~ \sqrt{|G|} e^{i \vec{x} \cdot \vec{k}} \langle \vec{k} | \psi \rangle.
  \label{<+label+>}
\end{equation}

\section{Conserved current in minimal length models}
\label{apx:current_MLQM}

Here, we will derive the form of the conserved current for a generic transformation starting from a Lagrangian depending on a field $\Psi$, its time derivative $\partial_0 \Psi$, and spatial derivatives of any order.

Let us consider a coordinate transformation of the form
\begin{equation}
  x^\mu \to x'{}^\mu = x^\mu - \delta x^\mu
  \label{<+label+>}
\end{equation}
and the corresponding field transformation
\begin{equation}
  \Psi \to \Psi' = \Psi + \delta \Psi.
  \label{<+label+>}
\end{equation}
Following standard procedures, e.g. as in \cite{Dickbook}, and assuming that the action is invariant under the transformation, i.e., the transformation acts as a symmetry, we are interested in the variation of the volume form
\begin{equation}
  \delta (\diff x \mathcal{L})
  = (\delta \diff x) \mathcal{L} + \diff x \delta \mathcal{L}
  = - \diff x (\partial_\mu \delta x^\mu) \mathcal{L} + \diff x \delta \mathcal{L}.
  \label{eqn:variation_volume_form}
\end{equation}
We will first focus on the variation of the Lagrangian.
We find
\begin{equation}
  \delta \mathcal{L}
  = - \delta x^\mu \partial_\mu \mathcal{L} 
  + \sum_{j=0}^{\infty} \frac{\partial \mathcal{L}}{\partial \Psi_{\mu_1\ldots\mu_j}} \partial_{\mu_1\ldots\mu_j} \left[\delta \Psi + (\delta x^\mu \partial_\mu \Psi)\right]
  + \text{similar terms for } \Psi^*.
\end{equation}
By induction, one can easily prove that
\begin{multline}
  \frac{\partial \mathcal{L}}{\partial \Psi_{\mu_1\ldots\mu_j}} \partial_{\mu_1\ldots\mu_j} \left[\delta \Psi + (\delta x^\mu \partial_\mu \Psi)\right]\\
  = \sum_{l=0}^{j} (-1)^{j-l} \binom{j}{l} 
  \partial_{\mu_{1}\ldots\mu_{l}} \left[ \left( \partial_{\mu_{l+1}\ldots\mu_j} \frac{\partial \mathcal{L}}{\partial \Psi_{\mu_1\ldots\mu_j}} \right) \left(\delta \Psi + \delta x^\mu \partial_\mu \Psi\right) \right].
\end{multline}
In the expression above, the derivative indices are enumerated in increasing order, \emph{i.e.}, $1 \leq \cdots \leq l < l+1 \leq \cdots \leq j$ and, \emph{e.g.}, the symbol $\partial_{\mu_{1}\ldots\mu_{l}}$ amounts to no derivative when $l < 1$.
Using Eq. \eqref{eqn:variation_volume_form} and since the action is invariant under the symmetry transformation, we find the off-shell relation
\begin{multline}
  \partial_{\mu_1} \left\{ \sum_{j=1}^{\infty} \sum_{l=1}^{j} (-1)^{j-l} \binom{j}{l}
  \partial_{\mu_2\ldots\mu_{l}} \left[ \left( \partial_{\mu_{l+1}\ldots\mu_j} \frac{\partial \mathcal{L}}{\partial \Psi_{\mu_1\ldots\mu_j}} \right) \left( \delta \Psi + \delta x^\mu \partial_\mu \Psi \right) \right] \right.\\
  \left. - \delta_\mu^{\mu_1} \delta x^\mu \mathcal{L} \right\} 
  + \text{similar terms for } \Psi^* \\
  = - \sum_{j=0}^{\infty} (-1)^j \left( \partial_{\mu_1\ldots\mu_j} \frac{\partial \mathcal{L}}{\partial \Psi_{\mu_1\ldots\mu_j}} \right) \left( \delta \Psi + \delta x^\mu \partial_\mu \Psi \right)
  + \text{similar terms for } \Psi^*.
  \label{<+label+>}
\end{multline}
Thus, on-shell, using the equations of motion for $\Psi$ in Eq. \eqref{eqn:eom_Psi} and the analogous for $\Psi^*$, we can identify a conserved current
\begin{multline}
  J^{\mu_1} 
  = \delta_\mu^{\mu_1} \delta x^\mu \mathcal{L}
  - \sum_{j=1}^{\infty} \sum_{l=1}^{j} (-1)^{j-l} \binom{j}{l} 
  \partial_{\mu_2\ldots\mu_{l}} \left[ \left( \partial_{\mu_{l+1}\ldots\mu_j} \frac{\partial \mathcal{L}}{\partial \Psi_{\mu_1\ldots\mu_j}} \right) \left( \delta \Psi + \delta x^\mu \partial_\mu \Psi \right) \right] \\
  + \text{similar terms for } \Psi^*.
  \label{eqn:GUP_current}
\end{multline}
It is convenient writing the same current as
\begin{multline}
  J^{\mu_1} 
  = \delta_\mu^{\mu_1} \delta x^\mu \mathcal{L}
  + \sum_{j=1}^{\infty} \sum_{l=1}^{j} (-1)^l \left( \partial_{\mu_2 \ldots \mu_l} \frac{\partial \mathcal{L}}{\partial \Psi_{\mu_1 \ldots \mu_j}} \right) 
  \partial_{\mu_{l+1} \ldots \mu_j} \left( \delta \Psi + \delta x^\mu \partial_\mu \Psi \right) \\
  + \text{similar terms for } \Psi^*.
  \label{eqn:GUP_current}
\end{multline}

Finally, we can introduce the energy-momentum tensor $\Theta_\mu{}^{\mu_1}$ for a constant coordinate shift as
\begin{equation}
  J^{\mu_1} = \Theta_\mu{}^{\mu_1} \delta x^\mu.
  \label{<+label+>}
\end{equation}
In this case, the fields transform like scalars obtaining
\begin{multline}
  \Theta_\mu{}^{\mu_1}
  = \delta^{\mu_1}_\mu \mathcal{L} 
  + \sum_{j=1}^{\infty} \sum_{l=1}^{j} (-1)^l \left[ \left(\partial_{\mu_2\ldots\mu_l} \frac{\partial \mathcal{L}}{\partial \Psi_{\mu_1\ldots\mu_j}}\right) \left( \partial_{\mu_{l+1}\ldots\mu_j \mu} \Psi \right) \right.\\
  \left. + \left( \partial_{\mu_2\ldots\mu_l} \frac{\partial \mathcal{L}}{\partial \Psi^*_{\mu_1\ldots\mu_j}} \right) \left( \partial_{\mu_{l+1}\ldots\mu_j \mu} \Psi^* \right) \right].
  \label{eqn:energy_momentum}
\end{multline}

\section{Current density in one dimension}
\label{apx:1D_current}

We will now adapt the results presented above to a one-dimensional model.
In this case, the free Lagrangian acquires the following form
\begin{equation}
  \mathcal{L}
  = \frac{i \hbar}{2} \left(\Psi^{*} \partial_0 \Psi - \partial_0 \Psi^{*} \Psi\right) 
  - \frac{\hbar^2}{2 m} g(- \partial^2) \partial \Psi^{*} g(- \partial^2) \partial \Psi.
\end{equation}
Thus, for the probability current, we get
\begin{multline}
  J 
  = \frac{i \hbar}{2 m} \sum_{N=0}^{\infty} \frac{(-1)^N}{N!} 
  \sum_{n=0}^{N} \binom{N}{n} \left[ \frac{\diff^{n}}{\diff x^{n}} g(k^2) \right]_{x=0} \left[ \frac{\diff^{N-n}}{\diff x^{N-n}} g(k^2) \right]_{x=0} \\
  \times \sum_{l=0}^{2n} (-1)^{l} \left[ \left( \partial^{(2N-l+1)} \Psi^{*} \right) \left( \partial^{(l)} \Psi \right) 
  - \left( \partial^{(2N-l+1)} \Psi \right) \left( \partial^{(l)} \Psi^* \right) \right].
\end{multline}
Let us momentarily consider the case with an odd value of $N$.
Since the factor
  \begin{equation}
    \binom{N}{n} \left[ \frac{\diff^{n}}{\diff x^{n}} g(k^2) \right]_{x=0} \left[ \frac{\diff^{N-n}}{\diff x^{N-n}} g(k^2) \right]_{x=0}
    \label{<+label+>}
  \end{equation}
  is invariant when substituting $n$ with $N-n$, we can write
  \begin{multline}
    \sum_{n=0}^{N} \binom{N}{n} \left[ \frac{\diff^{n}}{\diff x^{n}} g(k^2) \right]_{x=0} \left[ \frac{\diff^{N-n}}{\diff x^{N-n}} g(k^2) \right]_{x=0}\\
    \times \sum_{l=0}^{2n} (-1)^{l} \left[ \left( \partial^{(2N-l+1)} \Psi^{*} \right) \left( \partial^{(l)} \Psi \right) - \left( \partial^{(2N-l+1)} \Psi \right) \left( \partial^{(l)} \Psi^* \right) \right]\\
    = \sum_{n=0}^{\lfloor N/2 \rfloor} \binom{N}{n} \left[ \frac{\diff^{n}}{\diff x^{n}} g(k^2) \right]_{x=0} \left[ \frac{\diff^{N-n}}{\diff x^{N-n}} g(k^2) \right]_{x=0}\\
    \times \left\{\sum_{l=0}^{2n} (-1)^{l} \left[ \left( \partial^{(2N-l+1)} \Psi^{*} \right) \left( \partial^{(l)} \Psi \right) - \left( \partial^{(2N-l+1)} \Psi \right) \left( \partial^{(l)} \Psi^* \right) \right]\right.\\
    + \left. \sum_{l=0}^{2(N-n)} (-1)^{l} \left[ \left( \partial^{(2N-l+1)} \Psi^{*} \right) \left( \partial^{(l)} \Psi \right) - \left( \partial^{(2N-l+1)} \Psi \right) \left( \partial^{(l)} \Psi^* \right) \right]\right\}.
    \label{eqn:step1_oddN}
  \end{multline}
  Let us focus on the sum over $l$.
  We then have
  \begin{multline}
    \sum_{l=0}^{2(N-n)} (-1)^{l} \left[ \left( \partial^{(2N-l+1)} \Psi^{*} \right) \left( \partial^{(l)} \Psi \right) - \left( \partial^{(2N-l+1)} \Psi \right) \left( \partial^{(l)} \Psi^* \right) \right]\\
    = 2 \sum_{l=2n+1}^{N} (-1)^{l} \left[ \left( \partial^{(2N-l+1)} \Psi^{*} \right) \left( \partial^{(l)} \Psi \right) - \left( \partial^{(2N-l+1)} \Psi \right) \left( \partial^{(l)} \Psi^* \right) \right]\\
    + \sum_{l=0}^{2n} (-1)^{l} \left[ \left( \partial^{(2N-l+1)} \Psi^{*} \right) \left( \partial^{(l)} \Psi \right) - \left( \partial^{(2N-l+1)} \Psi \right) \left( \partial^{(l)} \Psi^* \right) \right].
  \end{multline}
  From this, we can write
  \begin{multline}
    \sum_{l=0}^{2(N-n)} (-1)^{l} \left[ \left( \partial^{(2N-l+1)} \Psi^{*} \right) \left( \partial^{(l)} \Psi \right) - \left( \partial^{(2N-l+1)} \Psi \right) \left( \partial^{(l)} \Psi^* \right) \right]\\
    + \sum_{l=0}^{2n} (-1)^{l} \left[ \left( \partial^{(2N-l+1)} \Psi^{*} \right) \left( \partial^{(l)} \Psi \right) - \left( \partial^{(2N-l+1)} \Psi \right) \left( \partial^{(l)} \Psi^* \right) \right]\\
    = 2 \sum_{l=0}^{N} (-1)^{l} \left[ \left( \partial^{(2N-l+1)} \Psi^{*} \right) \left( \partial^{(l)} \Psi \right) - \left( \partial^{(2N-l+1)} \Psi \right) \left( \partial^{(l)} \Psi^* \right) \right].
  \end{multline}
  The last expression is trivially valid also in the case of an even value for $N$ and $n=N/2$.
  Thus, we have
  \begin{multline}
    \sum_{n=0}^{N} \binom{N}{n} \left[ \frac{\diff^{n}}{\diff x^{n}} g(k^2) \right]_{x=0} \left[ \frac{\diff^{N-n}}{\diff x^{N-n}} g(k^2) \right]_{x=0}\\
    \times \sum_{l=0}^{2n} (-1)^{l} \left[ \left( \partial^{(2N-l+1)} \Psi^{*} \right) \left( \partial^{(l)} \Psi \right) - \left( \partial^{(2N-l+1)} \Psi \right) \left( \partial^{(l)} \Psi^* \right) \right]\\
    = \left[ \frac{\diff^{N}}{\diff x^{N}} g^2(k^2) \right]_{x=0} \sum_{l=0}^{2N+1} (-1)^{l} \left( \partial^{(2N-l+1)} \Psi^{*} \right) \left( \partial^{(l)} \Psi \right).
  \end{multline}
  When $N$ is even, Eq. \eqref{eqn:step1_oddN} becomes
  \begin{multline}
    \sum_{n=0}^{N} \binom{N}{n} \left[ \frac{\diff^{n}}{\diff x^{n}} g(k^2) \right]_{x=0} \left[ \frac{\diff^{N-n}}{\diff x^{N-n}} g(k^2) \right]_{x=0}\\
    \times \sum_{l=0}^{2n} (-1)^{l} \left[ \left( \partial^{(2N-l+1)} \Psi^{*} \right) \left( \partial^{(l)} \Psi \right) - \left( \partial^{(2N-l+1)} \Psi \right) \left( \partial^{(l)} \Psi^* \right) \right]\\
    = \left[ \frac{\diff^{N}}{\diff x^{N}} g^2(k^2) \right]_{x=0} \sum_{l=0}^{2N+1} (-1)^{l} \left( \partial^{(2N-l+1)} \Psi^{*} \right) \left( \partial^{(l)} \Psi \right).
  \end{multline}
  Thus, in both cases of even or odd $N$, we get
  \begin{equation}
    J 
    = \frac{i \hbar}{2 m} \sum_{N=0}^{\infty} \frac{(-1)^N}{N!} \left[ \frac{\diff^{N}}{\diff x^{N}} g^2(k^2) \right]_{x=0}
    \sum_{l=0}^{2N+1} (-1)^{l} \left( \partial^{(2N-l+1)} \Psi^{*} \right) \left( \partial^{(l)} \Psi \right).
  \end{equation}

\end{appendices}



\end{document}